\documentclass[11pt]{article}
\usepackage[pdftex]{graphicx}
\usepackage{booktabs}
\pdfoutput =1

\usepackage{array, amsmath, graphicx, mathtools, setspace, booktabs, float, caption, url}
\usepackage{palatino}
\usepackage[noblocks]{authblk}
\usepackage{subcaption, caption, abstract, txfonts}

\usepackage{cite}

\graphicspath{{Figures/}}

\usepackage[page]{appendix}
\usepackage{mathpazo, wrapfig, subcaption, tikz} 
\usepackage[scaled]{helvet} 
\usepackage{courier} 
\normalfont
\usepackage[T1]{fontenc}
\usetikzlibrary{%
   arrows,%
   calc,%
   fit,%
   patterns,%
   plotmarks,%
   shapes.geometric,%
   shapes.misc,%
   shapes.symbols,%
   shapes.arrows,%
   shapes.callouts,%
   shapes.multipart,%
   shapes.gates.logic.US,%
   shapes.gates.logic.IEC,%
   er,%
   automata,%
   backgrounds,%
   chains,%
   topaths,%
   trees,%
   petri,%
   mindmap,%
   matrix,%
   calendar,%
   folding,%
   fadings,%
   through,%
   positioning,%
   scopes,%
   decorations.fractals,%
   decorations.shapes,%
   decorations.text,%
   decorations.pathmorphing,%
   decorations.pathreplacing,%
   decorations.footprints,%
   decorations.markings,%
   shadows} 

\xdefinecolor{mygreen}{rgb}{0, 0, 1}
\xdefinecolor{myyell}{rgb}{1, 1, 0}
\xdefinecolor{mygreen}{rgb}{0.2, 0.8, 0.2}
\xdefinecolor{LightGreen}{rgb}{0.7, 1, 0.7}
\xdefinecolor{LightBlue}{rgb}{0.6, 0.6, 1}
\xdefinecolor{LightRed}{rgb}{ 1, 0.7, 0.7}
\xdefinecolor{LightGrey}{rgb}{0.7, 0.7, 0.7}
\xdefinecolor{LightYellow}{rgb}{1, 1, 0.7}
\xdefinecolor{Lav}{rgb}{.9, .9, 0.98}
\xdefinecolor{DarkV}{rgb}{.6,0,.8}

\xdefinecolor{DarkBlue}{rgb}{0,0,.45}

\providecommand{\keywords}[1]{\textbf{\textit{Keywords:}} #1}

\newcommand{\Imin}{$I_{\text{min }}$}
\newcommand{\Ip}{$I_{\text{proj }}$}

\newcommand{\Ib}{$I_{\text{broja }}$} 
\newcommand{\Ic}{ $I_{\text{ccs }}$}  

\newcommand{\Id}{$I_\text{dep }$ }

\begin{document}
\title{Contrasting information theoretic decompositions of modulatory and arithmetic interactions in neural information processing systems}
\author[1]{Jim W. Kay} 
\author[2]{William A. Phillips}
\affil[1]{Department of Statistics, University of Glasgow, UK, jim.kay@glasgow.ac.uk}
\affil[2]{Faculty of Natural Sciences, University of Stirling, UK, w.a.phillips@stirling.ac.uk}

\maketitle
\begin{abstract}
Biological and artificial neural systems are composed of many local processors, and their capabilities depend upon the transfer function that relates each local processor's outputs to its inputs. This paper uses a recent advance in the foundations of information theory to study the properties of local processors that use contextual input to amplify or attenuate transmission of information about their driving inputs. This advance enables the information transmitted by processors with two distinct inputs to be decomposed into those components unique to each input, that shared between the two inputs, and that which depends on both though it is in neither, i.e. synergy. The decompositions that we report here show that contextual modulation has information processing properties that contrast with those of all four simple arithmetic operators, that it can take various forms, and that the form used in our previous studies of artificial nets composed of local processors with both driving and contextual inputs is particularly well-suited to provide the distinctive capabilities of contextual modulation under a wide range of conditions. We argue that the decompositions reported here could be compared with those obtained from empirical neurobiological and psychophysical data under conditions thought to reflect contextual modulation. That would then shed new light on the underlying processes involved. Finally, we suggest that such decompositions could aid the design of context-sensitive machine learning algorithms.
\end{abstract}

\begin{flushleft}
\end{flushleft}
\section{Introduction}
Biological and artificial neural systems are composed of many local processors, such as neurons or microcircuits, that are interconnected to form various network architectures in which the connections are adaptable via various learning algorithms. The architectures of mammalian neocortex and of the artificial nets trained by deep learning have several hierarchical levels of abstraction. In mammalian neocortex there are also feedback and lateral connections within and between different levels of abstraction. The information processing capabilities of neural systems do not depend only on their learning algorithms and architectures, however. They also depend upon the transfer functions that relate each local processor's outputs to its inputs. These transfer functions are often described in terms of the simple arithmetic operators, e.g.~\cite{silver}. In biological systems feedback and lateral connections can modulate transmission of feedforward information in various ways. There is ample evidence that multiplicative and divisive gain modulation are widespread in mammalian neocortex; see reviews~\cite{Sal2, CH, ECN}, where there is also evidence for amplification and attenuation via contextual modulation~\cite{PCS}. A central goal of the work reported here is to compare this form of contextual modulation with additive, subtractive, multiplicative and divisive interactions. Artificial neural systems that use an analogous form of contextual modulation to amplify or attenuate feedforward transmission within hierarchical architectures have also been studied, e.g.~\cite{PKS, KFP, KP2}, but their computational potential has not yet been adequately explored. As part of a broad effort to explore the information processing capabilities of contextual modulation and apply it to real-world data-processing tasks, this paper shows how it is related to the arithmetic transfer functions. That may provide new insights into the biological data, and help us design data processing algorithms with some of the context-sensitivity of neocortex.

To this end we build upon recent extensions to the foundations of information theory, i.e. multivariate mutual information decomposition~\cite{ENT, KIDP}. This recent advance enriches our conception of `information processing' by showing how the mutual information between two inputs and a local processor's output can be decomposed into that unique to each of the two inputs, that shared with the two inputs, and that which depends upon both inputs, i.e. synergy~\cite{KIDP, WB, WLP, WPKLP}. Though decomposition is possible in principle when there are more than two inputs, the number of components rises more rapidly than exponentially as the number of inputs increase, limiting its applicability to cases where the contributions of only a few different input variables needs to be distinguished. This is not necessarily a severe limitation, however, because each input variable whose contribution to output is to be assessed can itself be computed from indefinitely many elementary input variables by either linear or non-linear functions. Here we show that decomposition in the case of only two integrated input variables is adequate to distinguish contextual modulation from the arithmetic operators.
\begin{figure}[H]
\centering
\includegraphics[scale=0.7]{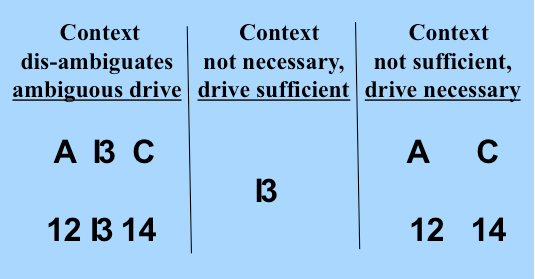}
\caption{An illustration of contextual modulation. \label{pict} }
\end{figure}

As the term `modulation' has been used to mean various different things it is important to note that the current paper is concerned with the conception of contextual modulation on which we have been working for many years~\cite{PCS, PKS, KFP, KP2}. It is similar to that used by some neurophysiologists, e.g.~\cite{Lamme},  and computational neuroscientists,  e.g.~\cite{Sal1},  and the present paper further develops that conception by using multivariate mutual information decomposition to contrast the effects of contextual modulation with those of the simple arithmetic operators. Thus, when in the ensuing we refer simply to `modulation' we mean contextual modulation, not multiplicative or divisive operations. A simple example of our notion of contextual modulation is shown in Fig.~\ref{pict}.

The driving information with which we are concerned here is that provided by the figure that can be seen as a B or as a 13. Each of these possible interpretations is at less than full strength because of the ambiguity. The two rows in the left column show that the probabilities of making each of the possible interpretations can be greatly influenced by the context in which the ambiguous figure is seen. The central column shows that drive alone is sufficient for some input information to be transmitted, and that context is not necessary because in its absence we can still see that a figure that could be a B or a 13 is present. The right column shows that drive is necessary and that context is not sufficient, because when the information that is dis-ambiguating in the left column is present in the absence of the driving information we do not see the B or 13 as being present.

This is but one example of indefinitely many, however, because from our perspective any signal can in principle  modulate the transmission of information about other signals. 
Though the information used for disambiguation in the above example comes from nearby locations in space, contextual modulation is not identified with any particular location in space because modulatory signals can in principle be from anywhere, including other points in time as well as other locations in space. Furthermore, though the example given here is contrived to make the ambiguity to be resolved obvious to introspection, the ambiguities that we assume to be resolvable by context are much more general, and include those due to weak signal to noise ratios, as well as those due to ambiguities of dynamic grouping and of task relevance. In short, we assume that contextual modulation uses other available information to amplify the transmission of `relevant' signals without corrupting their semantic content, i.e. what it is that they transmit information about. Multivariate mutual information transmission shows rigorously how that is possible.

The information whose transmission is amplified or attenuated can be referred to as the `driving' input, the `receptive field (RF)' input, or the input that specifies the neuron's `selective sensitivity'. Though none of these terms is wholly satisfactory, we refer to it as the `driving' input, and issues raised by that terminology will be noted in the final discussion.
 
A major contrast between hierarchies of abstractions in neocortex and those in deep learning systems is that intermediate levels of abstraction in neocortex typically have their own intrinsic utility, in addition to being on a path to even higher abstractions. There are outputs to subcortical sites from all levels of neocortical hierarchies, not only from a `top' level. This reduces the extent to which intermediate levels can become un-interpretable elements of a `black box', which is often thought to be a major weakness of current forms of deep learning. Another major contrast is that contextual modulation in neocortex provided by lateral and feedback connections guides the current processing of all inputs; whereas in nets trained by deep learning back-propagation from higher levels is used only for training the system how to respond to later inputs.

The work reported here contrasts with the simplifying assumption that neural systems are composed of integrate-and-fire neurons because that assumption implies a single site of integration in the modulatory context, whereas we study the information processing capabilities of local processors whose outputs depend on two functionally distinct integrated input variables, the drive and the modulatory context. Multivariate mutual information decomposition has already been applied to this case~\cite{WPKLP, KIDP},  and there is evidence that modulatory inputs are integrated at the distal end of the apical dendrite of some classes of neocortical pyramidal cell~\cite{PLHS, WAP}. 

To study these issues formally we use multivariate mutual information decomposition~\cite{WB} to analyze the information transmitted by one binary output variable given two input variables, and we  consider five different types of partial information decomposition.   The partial information decompositions of four different forms of contextual modulation are studied. Those decompositions are compared with decompositions of additive, subtractive, multiplicative, and divisive transfer functions. Here we consider continuous inputs that are generated from multimodal  and unimodal bivariate Gaussian probability models. Eight  different forms of interaction are considered for each of four scenarios in which the signal strengths given to the two inputs are different. The scenarios are chosen to show the distinctive information processing properties of contextual modulation. These properties are as follows. First, information about the drive can be transmitted when contextual modulation is absent or very weak, and all of it is transmitted if the drive is strong enough. Second, no information is transmitted about the modulatory input when drive is absent or very weak whatever the strength of the modulation. Third, modulatory inputs can amplify or attenuate the transmission of driving information that is weak or ambiguous.

\section{Methods}

\subsection{Notation and Definitions}

In~\cite{KIDP}, a local processor having two binary inputs and a binary output was discussed. 
Here, the inputs are instead continuous and have an absolutely continuous probability density function (p.d.f.), defined on $\mathbb{R}^2$. We denote the inputs to the processor by  the continuous random variables $R$ and $C$, and the output by the binary random variable, $Y$, taking values in the set $\{-1, 1\}$. The conditional distribution of $Y$, given that $R =r, C=c$ is Bernoulli with conditional probability of the logistic form
\begin{equation}
\Pr(Y=1|R=r, C=c) = 1/(1 + \exp{(-T(r, c))},  \label{condprob}
\end{equation}
where $T(r, c)$ is a continuous function defined on $\mathbb{R}^2$.

We now define the standard information theoretic terms that are required in this work and based on results in~\cite{CT}. We denote by the function $H$ the usual Shannon entropy, and note that any term with zero probabilities makes no contribution to the sums involved.
The total mutual information that is shared by $Y$ and the pair $(R, C)$ is given by,
\begin{equation}
I(Y; R, C) = H(Y) + H(R, C) - H(Y, R, C). \label{totmi}
\end{equation}

The information that is shared between $Y$ and $R$ but not with $C$ is 
\begin{equation}
I(Y; R |C) = H(Y, C) + H(R, C) - H(C) - H(Y, R, C), \label{yx1Gx2}
\end{equation}
and the information that is shared between $Y$ and $C$ but not with $R$ is
\begin{equation}
I(Y; C |R) = H(Y, R) + H(R, C) - H(R) - H(Y, R, C). \label{yx2Gx1}
\end{equation}
The information shared between $Y$ and $R$ is
\begin{equation}
I(Y; R) = H(Y) + H(R) - H(Y, R)
\end{equation}
and between $Y$ and $C$ is 
\begin{equation}
I(Y; C) = H(Y) + H(C) - H(Y, C)
\end{equation}

\subsection{Partial Information Decompositions}
Partial Information Decomposition (PID) was introduced by Williams and Beer~\cite{WB}. It provides a framework under which the mutual information, shared between the inputs and the output of a system, can be decomposed into components which measure four different aspects of the information: the unique information that each input conveys about the output; the shared information that both inputs posses regarding the output; the information that the inputs in combination have about the output.

Here  there are two inputs, $R, C$, and one output, $Y$.
 The information decomposition can be expressed as~\cite{WPKLP}
$$ I(Y; R, C) = I_{unq}(Y; R|C) +  I_{unq}(Y; C|R) + I_{shd}(Y; R, C) + I_{syn}(Y; R, C).$$

Adapting the notation of~\cite{WPKLP} we express our joint input mutual information in four terms as follows:
\begin{center}
\begin{tabular}{p{0.3\textwidth} p{0.6\textwidth}}
$\text{Unq}R  \equiv  I_{unq}(Y; R|C)$ & denotes the unique information that $R$ conveys about $Y$;\\\\
$ \text{Unq}C  \equiv I_{unq}(Y; C| R)$ & is the unique information that $C$ conveys about $Y$;\\\\
 $\text{Shd} \equiv I_{shd}(Y; R, C)$ & gives the shared (or redundant) information that both $R$ and $C$  have about $Y$;\\\\
$\text{Syn} \equiv I_{syn}(Y; R, C)$ & is the synergy or information that the joint variable $(R, C)$ has about $Y$ that cannot be obtained by observing $R$ and $C$ separately. 
\end{tabular}
\end{center}

It is possible to make deductions about a PID by using the following four equations which give a link between the components of a PID and certain classical Shannon measures of mutual information. The following are in~\cite[eqs. 4, 5]{WPKLP} , with amended notation; see also~\cite{WB}.
\begin{align}
I(Y; R) &=  \text{Unq}R +  \text{Shd} \label{ux1red}\\
I(Y; C) &=  \text{Unq}C +    \text{Shd}, \label{ux2red} \\
I(Y; R| C) &= \text{Unq}R +  \text{Syn}, \label{ux1syn}\\
I(Y; C | R) & =  \text{Unq}C + \text{Syn}. \label{ux2syn}
\end{align}

We consider here five different information decompositions.  The five methods of decomposition are $I_{\text{min}}$ \cite{WB}, $I_{\text{proj}}$ \cite{HSP}, $I_{\text{broja}}$ \cite{BERT},  $I_{\text{ccs}}$ \cite{RI} and $I_{\text{dep}}$ \cite{JEC}.
Although there are clear conceptual differences between them, where they agree we can have some confidence we are accurately decomposing information as would be appropriate.

We also consider residual output entropy defined by $H(Y)_{\text{res}} = H(Y|R, C)$. In addition to the four PID components, this residual measure is useful in expressing the information that is in $Y$ but not in $R, C$. It is also worth noting  that these five terms add up to the output entropy $H(Y)$, and when plotted we refer to the decomposition as a spectrum.
\begin{equation}
H(Y) = \text{UnqR} + \text{UnqC} + \text{Shd} + \text{Syn} + H(Y)_{\text{res}}.  \label{total}
\end{equation}
When making comparisons between different systems it is sometimes necessary to normalise the five measures in~$\eqref{total}$ by 
dividing each term by their total, the output entropy, $H(Y)$. 

In this study, the PID component, Shd, has not been separated into source and mechanistic terms~\cite{HSP, PICA, KIDP} because not all of the five PIDs considered  include definitions regarding how to achieve this task.

\subsection{Transfer  Functions}
We consider eight different forms of transfer function when computing the conditional output probabilities in~$\eqref{condprob}$. These transfer functions provide different ways of combining the two inputs. The functions in~$\eqref{TFmod}$ define modulatory forms of interaction, whereas those in~$\eqref{TFarith}$ are arithmetic. One aim of this study is to compare the PIDs obtained when using the different functions.

 There are four modulatory transfer functions that are defined as follows.
\begin{equation} \label{TFmod}
\begin{split}
T_{M_1} (r , c) &= \tfrac{1}{2} r ( 1 + \exp{( r c)}) \\
T_{M_2} (r , c) &= r + r  c \\
T_{M_3} (r , c) &=  r ( 1 + \tanh{(r c)}) \\
T_{M_4} (r, c) &= r 2^{r c} 
\end{split}
\end{equation}

We also consider four transfer functions that correspond to  arithmetic interactions between the inputs. They are given by
\begin{equation}
T_A(r, c) = r + c, \quad T_S(r, c) = r - c, \quad T_P(r, c) = r c, \quad T_D(r, c) = r/c.   \label{TFarith}\end{equation}

\subsection{Different Signal Strength Scenarios}

The inputs to the processor, $R, C$ are composed as $R =s_1 X_1$ and $C =s_2 X_2$, where $X_1, X_2$ are continuous random variables that have mean values of $\pm 1$. Then the mean values of $R$ are $\pm s_1$ and the mean values of $C$ are $\pm s_2$. The signal strengths, $s_1, s_2$, are non-negative real parameters that characterise the strength of each input.

Four combinations of signal strengths are used in this study, as defined in the following table. 
\begin{center}
\begin{tabular}{ll} \toprule
 Scenario 1: Strong $R$, with $C$ near to 0 &  ($s_1 =10.0, s_2 =0.05$)\\
Scenario 2: $R$ near to zero, with  strong $C$ &  ($s_1 =0.05, s_2 =10.0$)\\
Scenario 3: Weak $R$, with $C$ near to 0 & ($s_1 =1.00,  s_2 =0.05$)\\
Scenario 4: Weak $R$, with moderate $C$   & ($s_1 =1.00, s_2 = 5.00$)\\
\bottomrule
\end{tabular}
\end{center}
They are chosen to test for the key properties of contextual modulation.  They are: 
\begin{enumerate}
\item[CM1:] The drive, $R$, is sufficient for the output to transmit information about $R$, but context, $C$, is not necessary.
\item[CM2:] The drive, $R$, is necessary for the output to transmit information about $R$, but context, $C$, is not sufficient.
\item[CM3:] The output can transmit unique information about the drive, $R$, but not about the context, $C$.
\item[CM4:] The context can strengthen the transmission of information about $R$ when $R$ is weak.
\end{enumerate}

\subsection{Bivariate Gaussian Mixture Model (BGM)}

In earlier work \cite{KIDP}, we considered units that are bipolar. To continue this tradition, we  require the marginal distributions for the integrated receptive and contextual fields $R$ and $C$ to be bimodal, with the distribution of $R$ having modes at $\pm \,s_1$ and the distribution of $C$ having modes at $\pm \,s_2$. We shall define a bivariate Gaussian mixture model for $(R, C)$ which has four modes -- at $(- s_1,  - s_2), (-s_1, s_2), (s_1, -s_2)$ and $(s_1, s_2)$. First we consider the bivariate Gaussian mixture model for $(X_1, X_2)$, having probability density function 
\begin{equation}  f(x_1, x_2| \mu_1, \mu_2, \mu_3, \mu_4, \Sigma) = \sum_{i=1}^4 \pi_i f_i(x_1, x_2| \mu_i, \Sigma), \label{MM} \end{equation}
where 
$$ \mu_1 = \begin{bmatrix}-1\\  -1 \end{bmatrix}, \, \mu_2 = \begin{bmatrix}-1 \\1 \end{bmatrix}, \, \mu_3 = \begin{bmatrix} 1\\ -1\end{bmatrix}, \, \mu_4 = \begin{bmatrix}1\\ 1\end{bmatrix}, \,\, \Sigma = \sigma^2 \begin{bmatrix} 1 & \rho \\ \rho & 1 \end{bmatrix} $$
and the $\{ \pi_i \}$ are the mixing proportions  that are non-negative and sum to unity. For simplicity, we have assumed that the bivariate Gaussian pdfs, which form the four components of the mixture, have the same covariance matrix and also that the variances of $X_1$ and $X_2$ are equal. We note, in particular, that the correlation between $X_1$ and $X_2$ is equal to $\rho$ in all four of the component distributions. 

We require to investigate, however, what the correlation of $X_1$ and $X_2$ is for the mixture distribution defined by $ f(x_1, x_2| \mu_1, \mu_2, \mu_3, \mu_4, \Sigma)$ and, in particular, we require to find a way of setting this correlation to take any desired value in the simulations. We proceed as follows. 

\noindent Consider the random vector $X$ given by
$$ X = \begin{bmatrix} X_1 \\ X_2 \end{bmatrix}$$ Then, with respect to the mixture model $\eqref{MM}$, $X$ has mean vector
$$ \mathbb{E} (X) = \sum_{i=1}^4 \pi_i \, \mathbb{E} (X | i) =  \sum_{i=1}^4 \pi_i \, \mu_i = \begin{bmatrix} -\pi_1 - \pi_2 + \pi_3 + \pi_4 \\ -\pi_1 + \pi_2 - \pi_3 + \pi_4 \end{bmatrix},$$ where we have used the law of iterated expectation.
We make the simplifying assumption that $\pi_4 = \pi_1$ and that $\pi_3 = \pi_2$ which results in both components of $\mathbb{E}(X)$ being equal to zero. This assumption also means that $\pi_1 + \pi_2 = \tfrac{1}{2}.$

We denote the covariance matrix of $X$ in the mixture model $\eqref{MM}$ by $K$. Then
$$ K = \mathbb{E}[(X-\mu)( X - \mu)^T] = \mathbb{E} (X X^T) - \mu \mu^T = \mathbb{E}(X X^T), $$ since $\mu$ is the zero vector. Now, again using the law of iterated expectation, we have that
$$   \mathbb{E}(X X^T)= \sum_{i=1}^4 \pi_i \, \mathbb{E}(X X^T | i) = \sum_{i=1}^4 \pi_i \, (\Sigma + \mu_i \mu_i^T)= \Sigma +\sum_{i=1}^4 \pi_i \, \mu_i \mu_i^T, $$
which simplifies to
$$ K = \mathbb{E}(X X^T) = \begin{bmatrix} \sigma^2 + 1 & \rho \sigma^2 + 2 \pi_1 - 2 \pi_2 \\ \rho \sigma^2 + 2 \pi_1 - 2 \pi_2 &  \sigma^2 + 1 \end{bmatrix}. $$

Therefore, in the mixture model $\eqref{MM}$, the Pearson correlation coefficient of $X_1$ and $X_2$ is
$$ \text{cor}(X_1, X_2) = \frac{\rho \sigma^2 + 2 \pi_1 - 2 \pi_2}{\sigma^2 + 1}.$$
Setting this expression to be equal to the desired value of the correlation, $d$, gives that
$$ \pi_1 - \pi_2 = \tfrac{1}{2} ( d - (\rho -d) \sigma^2). $$
Taking $\rho = d$ and using the equality $\pi_1 + \pi_2 = \tfrac{1}{2}$ gives that
$$ \pi_1 = (1 + d)/4 \equiv \lambda \quad \text{and} \quad \pi_2 = (1-d)/4 \equiv \mu. $$

Hence, in order to generate values of $(X_1, X_2)$ from the BGM model that have correlation $d$ we take
$$ \Sigma = \sigma^2 \begin{bmatrix} 1 & d \\ d & 1 \end{bmatrix}, \quad \pi_4 = \pi_1 =\lambda \quad \text{and} \quad  \pi_3= \pi_2 = \mu,$$ for some pre-selected value of $\sigma$.
\begin{figure}[H]
\begin{tabular}{cc}
\begin{subfigure}{0.5\textwidth}\centering\includegraphics[scale=0.8]{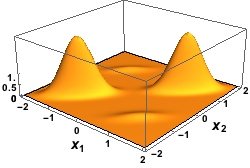}
\caption{$d =0.8$} \end{subfigure} &\begin{subfigure}{0.5\textwidth}\centering\includegraphics[scale=0.8]{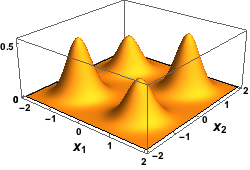}
\caption{$d=0.2$} \end{subfigure} \end{tabular}
\caption{ Probability density plots of $(X_1, X_2)$ in the BGM model for two different values of the correlation, $d$.  \label{bivG}}
\end{figure}

Now let $R = s_1 X_1$ and $C = s_2 X_2$. Then $(R, C)$ has the same correlation as $(X_1, X_2),$ by the invariance of Pearson correlation under linear transformation, and it has a bivariate Gaussian mixture model having component means
$$ \begin{bmatrix} -s_1 \\ -s_2 \end{bmatrix}, \,\, \begin{bmatrix} -s_1 \\ s_2 \end{bmatrix}, \,\, \begin{bmatrix} s_1 \\ -s_2 \end{bmatrix}, \,\,  \begin{bmatrix} s_1 \\ s_2 \end{bmatrix} $$
and component covariance matrices all equal to 
$$ \sigma^2 \begin{bmatrix} s_1^2 & d s_1 s_2 \\ d s_1 s_2 & s_2^2 \end{bmatrix}, $$ with mixing proportions $\{\lambda, \mu, \mu, \lambda\}$. It follows from the above discussion that in the bivariate Gaussian mixture model $\eqref{MM},$ $(R, C)$ has mean vector $\mu$ and covariance matrix $K$ given by
$$ \mu = \begin{bmatrix} 0 \\ 0 \end{bmatrix} \quad \text{and} \quad K = (\sigma^2 +1) \begin{bmatrix} 1 & d \\ d & 1 \end{bmatrix}. $$

Some plots of the BGM probability density function are given in Fig.~\ref{bivG}. To complete the 3D specification of the joint distribution for $(Y, R, C)$ the conditional distribution of $Y$, given that $R=r$ and $C =c$, was assumed to be Bernoulli with probability equal to the logistic function applied to $T(r, c)$, as in~$\eqref{condprob}$, and the full 3D model is termed the BGMB model.

\subsection{Single Bivariate Gaussian Model (SBG)}
While in our previous work~\cite{KIDP}, and in Section 2.5, bipolar inputs have been considered, there are many datasets in neuroscience and in machine learning where the input variables are not bimodal. We extend our approach, therefore, by considering the input data  to have been generated from a single bivariate Gaussian probability model. We take $(X_1, X_2)$ to have a bivariate Gaussian distribution, with mean vector and covariance matrix 
$$ \mu = \begin{bmatrix} 1 \\ 1 \end{bmatrix}, \quad \Sigma = \sigma^2\begin{bmatrix}  1 & d \\ d & 1   \end{bmatrix}. $$ Then, $R$ and $C$ are bivariate Gaussian, with mean vector and covariance matrix
$$ \mu = \begin{bmatrix} s_1 \\ s_2 \end{bmatrix}, \quad \Sigma = \sigma^2\begin{bmatrix}  s_1^2 & ds_1s_2 \\ ds_1s_2 & s_2^2   \end{bmatrix}. $$
Taking $\sigma =0.3$, ensures that almost 100\% of the generated values of $(X_1, X_2)$ lie within the unit square $[0, 2] \times [0, 2]$, and so the corresponding values of $R$ and $C$ are almost certainly positive. 
\newline
Given positive inputs, it is necessary to introduce a bias term, $b$,  into each transfer function. Thus, the general $T(r, c)$ is replaced here by
$$ T_b(r,c) = T(r, c) - b, $$
where $b$ is taken to be the median of the values of $T(r, c)$ obtained by plugging in the generated values of $r$ and $c$. This choice of $b$ ensures that the simulated outputs are equally likely to be $+1$ or $-1$, so the output entropy $H(Y)$ will be very close to its maximum value of 1. The conditional output probabilities are computed using~$\eqref{condprob}$ , with $T(r, c)$ replaced by $T_b(r, c)$  for each of the transfer functions in~$\eqref{TFmod}$-$\eqref{TFarith}$.

\subsection{The First Simulation}
The first simulation study has four factors: transfer function, PID method, scenario and the correlation between the inputs. For each combination of the eight transfer functions, four signal-strength scenarios and two values of correlation (0.8, 0.2), data were generated using the following procedure.

One million samples of $(X_1, X_2)$ were generated from the bivariate Gaussian mixture model in~$\eqref{MM}$ for a given value of the correlation, $d$, a given combination of the signal strengths, $s_1, s_2$ and a given transfer function.  In all cases the standard deviation of $X_1$ and $X_2$ was taken as $\sigma =0.3$. 

For each sample, the values of $R$ and $C$ were computed as $r =s_1 x_1$ and $c=s_2 x_2$, and these values were passed through the appropriate transfer function to compute the conditional output probability, $\theta_{r c}$ in~$\eqref{condprob}$. For each value of $r,c$ a value of $Y$ was simulated from the Bernoulli distribution having probability, $\theta_{rc}$, thus giving a simulated data set for $(Y, R, C)$.

 Given the lack of PIDs for this type of data, it is necessary to bin the data. For each data set, a hierarchical binning procedure was employed. The values of $R$ were separated according to sign, and then the positive values were split equally into three bins using tertiles, and similarly for the negative values of $R$. The same procedure was used to define bins for the values of $C$.

 The values of the binary output $Y$ were binned according to sign. Having defined the bins, the one million $(Y, R, C)$ observations were allocated to the appropriate bins. This procedure produced a $6 \times 6 \times 2$ probability array that was used in each case as input for the computation of each of the five PIDs used in the study, making use of the package, {\it dit}, which is available on GitHub\footnote{ \url{https://github.com/dit/dit}}. 
 
 It should be noted that for some data sets other sizes of probability array were considered, including $4 \times 4 \times 2$, $5 \times 5 \times 2$, $8 \times 8 \times 2$, and it was found that the results for the $6 \times 6 \times 2$ and $8 \times 8 \times 2$ arrays were very close, which suggests that using a $6 \times 6 \times 2$ array is sufficient.
 
\subsection{The Second Simulation}
The second simulation study is essentially a repeat of the first study, as described in Section 2.6, with the continuous input data for $R$ and $C$ being generated using the SBG model of Section 2.6 rather than the BGM model. The values of the binary output $Y$ were simulated in a similar way, with transfer functions of the form $T_b(r, c)$ used rather than $T(r, c)$.  The binning of the data was performed in a slightly different way. Given the absence of bipolarity here, the values of $R$ were divided equally into six bins defined by sextiles, and similarly for $C$. The binary outputs were split according to sign. Thus, again here, a $6 \times 6 \times 2$ probability array was created and it was used in each case to compute the PIDs.

\section{The Results}
The PIDs obtained using the methods, \Imin, \Ip, \Ib, were almost identical (some small differences in the fourth decimal place) in both simulations,  so we show results only for the \Ib method. Thus, anything said in the ensuing about this method also applies to the \Imin and \Ip PIDs.

\subsection{First simulation}

\begin{figure}[H]
 \begin{center}
 \includegraphics[height=.7cm]{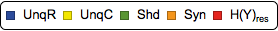}
 \end{center}
\begin{tabular}[t]{lccc}
& \Ib & \Id & \Ic \\
\begin{tabular}{l}
\small{$s_1$ = 10.0 }\\
\small{$s_2$ = 0.05} \\
\vspace{0.3cm}
\end{tabular} & \begin{subfigure}{0.3\textwidth}\centering\includegraphics[width=4.5cm]{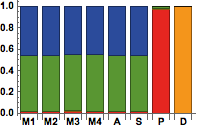}
\caption{}\end{subfigure} &\begin{subfigure}{0.3\textwidth}\centering\includegraphics[width=4.5cm]{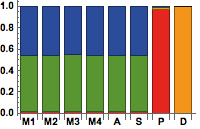}
 \caption{}\end{subfigure} & \begin{subfigure}{0.3\textwidth}\centering\includegraphics[width=4.5cm]{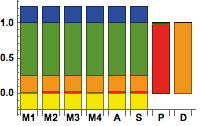}
\caption{} \end{subfigure}  \\
\begin{tabular}{l}
\small{$s_1$ =  0.05 }\\
\small{$s_2$ = 10.0} \\
\vspace{0.3cm}
\end{tabular} &\begin{subfigure}{0.3\textwidth}\centering\includegraphics[width=4.5cm]{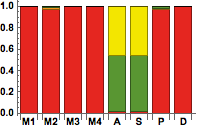}
\caption{}\end{subfigure} &\begin{subfigure}{0.3\textwidth}\centering\includegraphics[width=4.5cm]{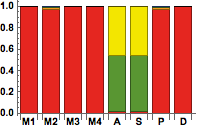}
 \caption{}\end{subfigure} & \begin{subfigure}{0.3\textwidth}\centering\includegraphics[width=4.5cm]{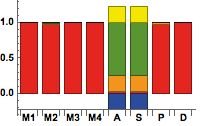}
\caption{} \end{subfigure}  \\
\begin{tabular}{l}
\small{$s_1$ = 1.0 }\\
\small{$s_2$ = 0.05} \\
\vspace{.3cm}
\end{tabular} &\begin{subfigure}{0.3\textwidth}\centering\includegraphics[width=4.5cm]{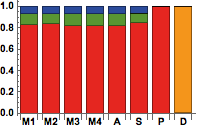}
\caption{}\end{subfigure} &\begin{subfigure}{0.3\textwidth}\centering\includegraphics[width=4.5cm]{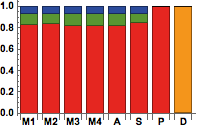}
 \caption{}\end{subfigure} & \begin{subfigure}{0.3\textwidth}\centering\includegraphics[width=4.5cm]{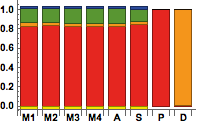}
\caption{} \end{subfigure}  \\
\begin{tabular}{l}
\small{$s_1$ = 1.0 }\\
\small{$s_2$ = 5.0} \\
\vspace{0.3cm}
\end{tabular} &\begin{subfigure}{0.3\textwidth}\centering\includegraphics[width=4.5cm]{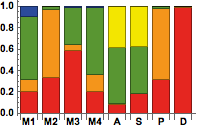}
\caption{}\end{subfigure} &\begin{subfigure}{0.3\textwidth}\centering\includegraphics[width=4.5cm]{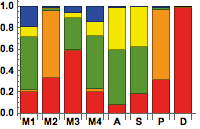}
 \caption{}\end{subfigure} & \begin{subfigure}{0.3\textwidth}\centering\includegraphics[width=4.5cm]{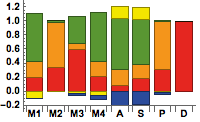}
\caption{} \end{subfigure}  \end{tabular}
\caption{Normalised spectra for the \Ib, \Id and  \Ic  PIDs, given four combinations of the signal strength parameters, $(s_1, s_2)$.  The data were generated from the BGM model. The correlation between inputs, $R$ and $C$,  is 0.8.  \label{corr.8}}
\end{figure}

Here, each input is generated from a bimodal Gaussian distribution which has modes centred on -1 and +1. The results are presented in Figs.~\ref{corr.8}-\ref{corr.2} in a way that makes it easy to make many different comparisons if the color scheme for depicting the five different components is learned. They show that different transfer functions can transmit similar components of input information in some signal-strength scenarios, but very different components under other scenarios. The rich variety of spectra produced provides a new perspective on the information processing capabilities of each of the transfer functions. In particular, they show that the components of information transmitted by contextual modulation differ greatly from those of all four simple arithmetic operators, including multiplication and division, which are also often thought of as modulatory.

\setlength{\tabcolsep}{0.01pt}
\setlength{\tabcolsep}{6pt}

\subsubsection{Distinctive information transmission properties of contextual modulation}
The distinctive properties of contextual modulation can be seen by comparing the \Ib decomposition of M1, i.e. the left-most spectra in column 1 of Fig.~\ref{corr.8}, with those of all four simple arithmetic operators. The four rows show these decompositions for each of the four different signal-strength scenarios. M1 transmits the same or very similar components as additive and subtractive operators when context strength is close to zero. Thus, in the absence of contextual information, it defaults to near-equivalence with the additive and subtractive operators. The effects of contextual modulation differ greatly from those of addition and subtraction when drive strength approaches zero, however. In that case no input information is transmitted by the modulatory interaction whereas additive and subtractive operators transmit all or most of the input information in the form of the shared component and that unique to the  stronger input. These comparisons show that drive is both necessary and sufficient for transmission of input information by the contextual modulatory interaction, M1, and that contextual input is neither necessary nor sufficient. Thus, this clearly displays the marked asymmetry between the effects of driving and modulatory input that is a distinctive property of contextual modulation.
\setlength{\tabcolsep}{0.01pt}
\begin{figure}[H]
 \begin{center}
 \includegraphics[height=.7cm]{legend.png}
 \end{center}
\begin{tabular}[t]{lccc}
& \Ib & \Id & \Ic \\
\begin{tabular}{l}
\small{$s_1$ = 10.0 }\\
\small{$s_2$ = 0.05} \\
\vspace{0.3cm}
\end{tabular} & \begin{subfigure}{0.3\textwidth}\centering\includegraphics[width=4.5cm]{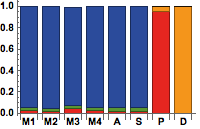}
\caption{}\end{subfigure} &\begin{subfigure}{0.3\textwidth}\centering\includegraphics[width=4.5cm]{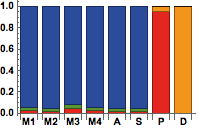}
 \caption{}\end{subfigure} & \begin{subfigure}{0.3\textwidth}\centering\includegraphics[width=4.5cm]{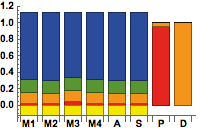}
\caption{} \end{subfigure}  \\
\begin{tabular}{l}
\small{$s_1$ =  0.05 }\\
\small{$s_2$ = 10.0} \\
\vspace{0.3cm}
\end{tabular} &\begin{subfigure}{0.3\textwidth}\centering\includegraphics[width=4.5cm]{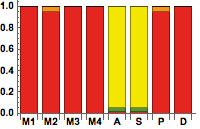}
\caption{}\end{subfigure} &\begin{subfigure}{0.3\textwidth}\centering\includegraphics[width=4.5cm]{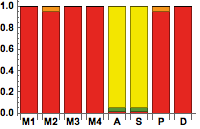}
 \caption{}\end{subfigure} & \begin{subfigure}{0.3\textwidth}\centering\includegraphics[width=4.5cm]{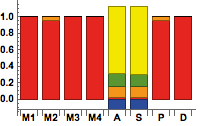}
\caption{} \end{subfigure}  \\
\begin{tabular}{l}
\small{$s_1$ = 1.0 }\\
\small{$s_2$ = 0.05} \\
\vspace{.3cm}
\end{tabular} &\begin{subfigure}{0.3\textwidth}\centering\includegraphics[width=4.5cm]{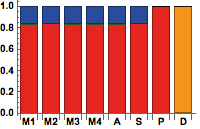}
\caption{}\end{subfigure} &\begin{subfigure}{0.3\textwidth}\centering\includegraphics[width=4.5cm]{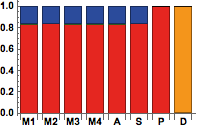}
 \caption{}\end{subfigure} & \begin{subfigure}{0.3\textwidth}\centering\includegraphics[width=4.5cm]{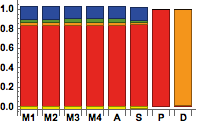}
\caption{} \end{subfigure}  \\
\begin{tabular}{l}
\small{$s_1$ = 1.0 }\\
\small{$s_2$ = 5.0} \\
\vspace{0.3cm}
\end{tabular} &\begin{subfigure}{0.3\textwidth}\centering\includegraphics[width=4.5cm]{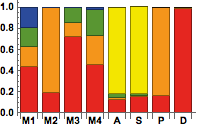}
\caption{}\end{subfigure} &\begin{subfigure}{0.3\textwidth}\centering\includegraphics[width=4.5cm]{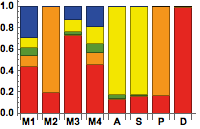}
 \caption{}\end{subfigure} & \begin{subfigure}{0.3\textwidth}\centering\includegraphics[width=4.5cm]{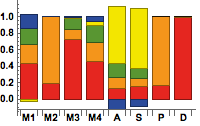}
\caption{} \end{subfigure}  \end{tabular}
\caption{Normalised spectra for the \Ib, \Id and  \Ic  PIDs, given four combinations of the signal strength parameters, $(s_1, s_2)$.   The data were generated from the BGM model. The correlation between inputs, $R$ and $C$,  is 0.2.  \label{corr.2}}
\end{figure}
\setlength{\tabcolsep}{6pt}

The components transmitted by M1 differ greatly from those of the multiplicative and divisive operators under all signal-strength scenarios. The multiplicative and divisive operators 
predominantly transmit synergistic components. They never transmit unique components. Contextual modulation does transmit some synergy, but that is much smaller than that transmitted by either multiplication or division, and occurs only when drive is weak. When drive is strong enough it transmits large components of both shared information and of that unique to the drive.

Though contextual modulation is neither necessary nor sufficient for information transmission, its modulatory effects can be clearly seen by comparing the components transmitted by M1 when drive is weak and context is moderate in Fig.~\ref{corr.8}(j), with those transmitted when drive is weak and context is near zero in Fig.~\ref{corr.8}(g). Moderate context increases the transmission of information about the drive by 0.4 bits. Stronger contextual inputs increase it even further. This increase is predominantly in the shared information but does include some increase in the component unique to the drive.

These fundamental contrasts between contextual modulation and the arithmetic operators apply also to the forms of interaction labeled as M2, M3, and M4, but with a few differences in some of the signal-strength scenarios. No differences between the four forms of contextual modulation occur when the strength of either the drive or the context is near zero. When neither is near zero the main effect of M2 is to greatly increase the synergistic component, resembling P. When drive is weak and context is strong M2 transmits mainly synergy, and is not at all effective in amplifying transmission of either the unique or shared information in the drive. M3 and M4 in that case do amplify transmission of shared information, but not of that unique to the drive.  Over all scenarios only M1 uses context to increase the transmission of information unique to the drive. 

In Fig.~\ref{corr.2}, the correlation between $R$ and $C$ is weak (0.2), rather than strong (0.8), and this has led to some clear differences in the expression of some information components when the spectra are compared with the corresponding spectra in Fig.~\ref{corr.8}. For all transfer functions, except P and D, there has been a large reduction in shared information, with a corresponding increase in the values of the unique informations. This is not surprising since the part of the shared information that is due to the correlation between the inputs must decrease as the correlation between inputs is changed from strong to weak. 
\newline
All of the messages regarding the comparisons of the modulatory with the arithmetic transfer functions that are discussed above, where the correlation between inputs is strong, apply here also when this correlation is weak, as do the conclusions regarding contextual modulation.

\subsubsection{Comparison of five different forms of information decomposition}
Encouraging convergence of the PID methodologies is indicating by our finding that the \Ib    \linebreak decomposition gives essentially the same results as \Imin and \Ip. Furthermore, the distinctive properties of contextual modulation seen when using the \Ib decomposition still apply when using the other forms, except for a few differences that are as follows. \Id  differs from \Ib, \Imin and \Ip, mainly in that when neither drive nor context are near zero it shows that M1, M3 and M4 transmit some information unique to the context, whereas \Ib does not. Therefore, the  \Id PIDs in Fig.~\ref{corr.8}(k) do not support condition CM3 for contextual modulation.

In contrast to all the other methods the \Ic PID can have negative unique information components. From~$\eqref{ux1red}$-$\eqref{ux2syn}$ it can be seen that a negative value for UnqR or UnqC corresponds the value of Shd or Syn or both being larger than they would be if the unique components were non-negative. Therefore, the \Ic spectra can cover a range of
approximately 1.4 instead of the range of 1 for the other PID methods. Nevertheless, in Scenario 1, the spectra for P and D are exactly the same as those obtained using \Ib and \Id. For the other transfer functions, which have almost identical spectra, the fact that UnqC is about -0.2 means that the synergy and shared components in these six spectra are correspondingly larger by 0.2. In Scenario 2, all the spectra are the same except for A and S. In this case, a negative value for UnqR leads to the presence of non-zero synergy and a larger estimate of the shared information. The spectra for Scenario 3 are almost the same as the corresponding transfer functions with \Ib and \Id, and so similar comments apply. In Scenario 4, the patterns of the spectra and the comparisons between them are essentially the same as those in the \Ib PIDs and so similar comparisons apply here, although the presence of negative unique components distorts the values of the other PID components, thus giving the spectra a somewhat different appearance.

In all four scenarios, transfer functions P and D have the same spectra for all five of the PID methods. For scenarios 1-3, all five PID methods show the same results for each transfer function, with the exception of functions A and S in Scenario 2 (d, e, f) in which \Ic has different spectra than those obtained with the other four methods. In Scenario 4, the spectra produced using \Id are very similar to those obtained with \Imin, \Ip and \Ib except for the presence of larger values of UnqR and UnqC.  Apart from functions P and D, the \Ic spectra appear to be rather different from the corresponding spectra given with the other four methods due to the presence of negative values for UnqR and/or UnqC.

When the correlation between inputs is weak, as in Fig.~\ref{corr.2}, all five PID methods produce the same spectra for P and D. For the remaining transfer functions, the corresponding  \Ib and \Id spectra are identical when either the drive or context is close to zero. When the drive is weak and the context is moderate, the corresponding \Id and \Ib spectra are virtually identical for the arithmetic transfer functions, along with M2, whereas for M1, M2 and M3 \Id produces larger unique components, and in particular positive values of information unique to the context. 
\newline
When the correlation between inputs is weak, the negative components in  \Ic are smaller in magnitude. Apart from the negative unique components, the \Ic spectra are very similar to those of the \Ib and \Id methods when either drive or context is close to zero. When neither is close to zero, the \Ic spectra are more like the corresponding \Ib spectra than those due to \Id because \Id has larger values for information unique to the context, $C$.
\newline
The same messages regarding contextual modulation and the comparison of PID methods, that are given above  in the case of strong correlation between inputs, apply here also when that correlation is weak.

Despite these differences the overall picture that emerges is that the distinctive properties of contextual modulation and their fundamental contrasts with the four simple arithmetic operators can be clearly seen in the decompositions produced by all five methods for computing the decompositions. The only exception with regard to contextual modulation is that noted for $I_{\text{dep}}$.  Whether the differences noted between the various methods have any functional significance remains to be determined.

\subsection{Second simulation}
\subsubsection{Distinctive information transmission properties of contextual modulation}
\begin{figure}[H]
 \begin{center}
 \includegraphics[height=.7cm]{legend.png}
 \end{center}
\begin{tabular}[t]{lccc}
& \Ib & \Id & \Ic \\
\begin{tabular}{l}
\small{$s_1$ = 10.0 }\\
\small{$s_2$ = 0.05} \\
\vspace{0.3cm}
\end{tabular} & \begin{subfigure}{0.3\textwidth}\centering\includegraphics[width=4.5cm]{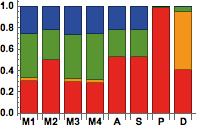}
\caption{}\end{subfigure} &\begin{subfigure}{0.3\textwidth}\centering\includegraphics[width=4.5cm]{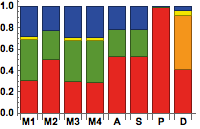}
 \caption{}\end{subfigure} & \begin{subfigure}{0.3\textwidth}\centering\includegraphics[width=4.5cm]{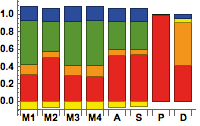}
\caption{} \end{subfigure}  \\
\begin{tabular}{l}
\small{$s_1$ =  0.05 }\\
\small{$s_2$ = 10.0} \\
\vspace{0.3cm}
\end{tabular} &\begin{subfigure}{0.3\textwidth}\centering\includegraphics[width=4.5cm]{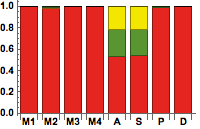}
\caption{}\end{subfigure} &\begin{subfigure}{0.3\textwidth}\centering\includegraphics[width=4.5cm]{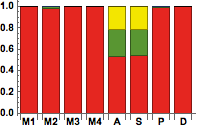}
 \caption{}\end{subfigure} & \begin{subfigure}{0.3\textwidth}\centering\includegraphics[width=4.5cm]{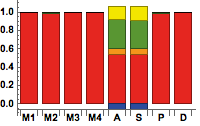}
\caption{} \end{subfigure}  \\
\begin{tabular}{l}
\small{$s_1$ = 1.0 }\\
\small{$s_2$ = 0.05} \\
\vspace{.3cm}
\end{tabular} &\begin{subfigure}{0.3\textwidth}\centering\includegraphics[width=4.5cm]{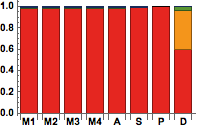}
\caption{}\end{subfigure} &\begin{subfigure}{0.3\textwidth}\centering\includegraphics[width=4.5cm]{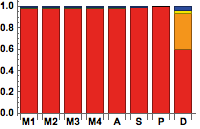}
 \caption{}\end{subfigure} & \begin{subfigure}{0.3\textwidth}\centering\includegraphics[width=4.5cm]{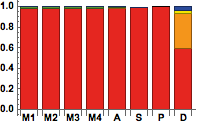}
\caption{} \end{subfigure}  \\
\begin{tabular}{l}
\small{$s_1$ = 1.0 }\\
\small{$s_2$ = 5.0} \\
\vspace{0.3cm}
\end{tabular} &\begin{subfigure}{0.3\textwidth}\centering\includegraphics[width=4.5cm]{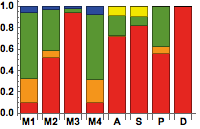}
\caption{}\end{subfigure} &\begin{subfigure}{0.3\textwidth}\centering\includegraphics[width=4.5cm]{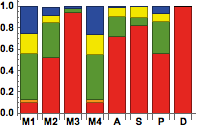}
 \caption{}\end{subfigure} & \begin{subfigure}{0.3\textwidth}\centering\includegraphics[width=4.5cm]{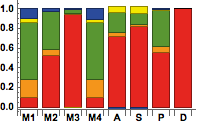}
\caption{} \end{subfigure}  \end{tabular}
\caption{Spectra in absolute units (bits) for the \Ib, \Id and  \Ic  PIDs, given four combinations of the signal strength parameters, $(s_1, s_2)$.  The data were generated from the SBG model.  The correlation between inputs, $R$ and $C$,  is 0.8.  \label{SGcorr.8}}
\end{figure}

In this simulation, each input is generated from a unimodal Gaussian distribution and is almost certainly positive. The spectra for the case of strong correlation between inputs are in Fig.~\ref{SGcorr.8}. As with the spectra in Section 3.1, we leave aside discussion of transfer functions P and D. When either  the drive or context is close to zero, the modulatory transfer functions indicate that conditions CM1 and CM2 are satisfied. 
 \setlength{\tabcolsep}{0.01pt}
\setlength{\tabcolsep}{6pt}

Furthermore, comparison of the corresponding spectra between scenarios 3 and 4 shows  that conditions CM3 and CM4 also hold, although only marginally for M3; M1, M2 and M4 transmit mainly shared information along with some synergy and information unique to the drive.  Transfer functions A and S are different from M1-M4 in that they express information that is unique to whichever is the stronger input, and they have non-zero values of UnqC in scenarios 2 and 4, thus breaking condition CM3.

When either the drive or context is close to zero P transmits virtually no information; when this is not the case, P expresses mainly shared information, and has a similar spectra to those of M2 for all five PIDs. Transfer function D transmits virtually no information in scenarios 2 and 4 where the context is moderate or strong, but when the context is close to zero it transmits mainly synergistic information.

\begin{figure}[H]
 \begin{center}
 \includegraphics[height=.7cm]{legend.png}
 \end{center}
\begin{tabular}[t]{lccc}
& \Ib & \Id & \Ic \\
\begin{tabular}{l}
\small{$s_1$ = 10.0 }\\
\small{$s_2$ = 0.05} \\
\vspace{0.3cm}
\end{tabular} & \begin{subfigure}{0.3\textwidth}\centering\includegraphics[width=4.5cm]{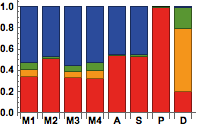}
\caption{}\end{subfigure} &\begin{subfigure}{0.3\textwidth}\centering\includegraphics[width=4.5cm]{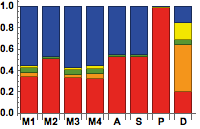}
 \caption{}\end{subfigure} & \begin{subfigure}{0.3\textwidth}\centering\includegraphics[width=4.5cm]{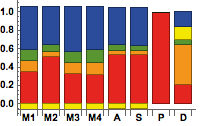}
\caption{} \end{subfigure}  \\
\begin{tabular}{l}
\small{$s_1$ =  0.05 }\\
\small{$s_2$ = 10.0} \\
\vspace{0.3cm}
\end{tabular} &\begin{subfigure}{0.3\textwidth}\centering\includegraphics[width=4.5cm]{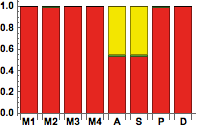}
\caption{}\end{subfigure} &\begin{subfigure}{0.3\textwidth}\centering\includegraphics[width=4.5cm]{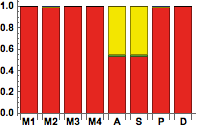}
 \caption{}\end{subfigure} & \begin{subfigure}{0.3\textwidth}\centering\includegraphics[width=4.5cm]{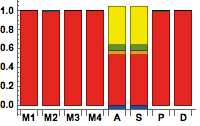}
\caption{} \end{subfigure}  \\
\begin{tabular}{l}
\small{$s_1$ = 1.0 }\\
\small{$s_2$ = 0.05} \\
\vspace{.3cm}
\end{tabular} &\begin{subfigure}{0.3\textwidth}\centering\includegraphics[width=4.5cm]{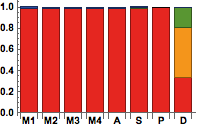}
\caption{}\end{subfigure} &\begin{subfigure}{0.3\textwidth}\centering\includegraphics[width=4.5cm]{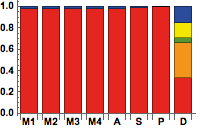}
 \caption{}\end{subfigure} & \begin{subfigure}{0.3\textwidth}\centering\includegraphics[width=4.5cm]{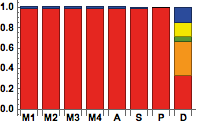}
\caption{} \end{subfigure}  \\
\begin{tabular}{l}
\small{$s_1$ = 1.0 }\\
\small{$s_2$ = 5.0} \\
\vspace{0.3cm}
\end{tabular} &\begin{subfigure}{0.3\textwidth}\centering\includegraphics[width=4.5cm]{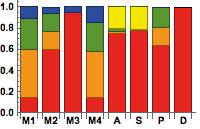}
\caption{}\end{subfigure} &\begin{subfigure}{0.3\textwidth}\centering\includegraphics[width=4.5cm]{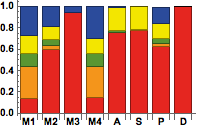}
 \caption{}\end{subfigure} & \begin{subfigure}{0.3\textwidth}\centering\includegraphics[width=4.5cm]{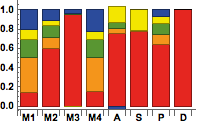}
\caption{} \end{subfigure}  \end{tabular}
\caption{Spectra in absolute units (bits) for the \Ib, \Id and  \Ic  PIDs, given four combinations of the signal strength parameters, $(s_1, s_2)$.  The data were generated from the SBG model.  The correlation between inputs, $R$ and $C$,  is 0.2.  \label{SGcorr.2}}
\end{figure}

When the correlation between inputs is weak, in Fig.~\ref{SGcorr.2}, there is generally less shared information and larger values for the unique information components and the synergistic information. The same points made in the case of strong correlation between inputs also apply here: M1, M2 and M4 possess the properties CM1-CM4 required for contextual modulation, and M4 marginally so. The arithmetic transfer functions do not have the modulation properties required.
 \setlength{\tabcolsep}{0.01pt}
\setlength{\tabcolsep}{6pt}
\subsubsection{Comparison of five different forms of information decomposition}
Ignoring the negative spectral components in the \Ic PIDs, all five PID methods produce very similar spectra in scenarios 1-3 for all transfer functions except D. Transfer function D gives the same spectra across all PID methods in Scenarios 2 and 4, but in scenarios 1 and 3 the other PID methods express the same level of synergistic information, but where \Ib gives a small shared component \Id and \Ic have small unique information components. 
\newline
In Scenario 4, where the drive is weak and the context is moderate, the corresponding \Ib and \Ic spectra are rather similar, although \Ic has some UnqC with M1 and M4. The corresponding \Id spectra for all transfer functions, except for M1, M2 and M4, has similar spectra to those given by \Ib and \Ic; for M1, M2 and M4, \Id has larger unique components, including quite large UnqC components with M1 and M4. 
\newline
Only the \Ib, \Imin and \Ip methods strictly possess all the requirements CM1-CM4 required to demonstrate contextual modulation, although the \Ic method almost does since its UnqC components are rather small.

\section{Conclusions and Discussion}
Multivariate mutual information decomposition shows that contextual modulation has properties that contrast with those of all four arithmetic operators, and that it can take various forms. They also raise several issues that need to be resolved.

\subsection{Contextual modulation contrasts with the arithmetic operators}
The information processing properties of contextual modulation are very different from those of the arithmetic operators. Though the modulatory transfer functions studied here are defined by the arithmetic operators their information processing properties are not in any simple sense a compilation of those of the individual operators used to define them. Outputs from additive and subtractive interactions transmit unique information about whichever input is strongest, whereas the modulatory functions transmit unique information only about the drive. Multiplicative and divisive operators also differ greatly from the contextual modulation studied above. In contrast to contextual modulation, outputs from multiplicative interactions predominantly transmit information about synergistic relations between the two inputs. It is very small if either of the inputs is near zero, and very near zero if both are near zero. Outputs from divisive interactions also predominantly transmit synergistic information. It approaches zero as the divisor increases in strength relative to the drive.

The contextual modulations studied here transmit information unique to the driving input without transmitting information little or no unique to the modulator though using it to amplify or attenuate transmission of information about the drive. When the driving input is very weak modulatory functions transmit little or no input information, whereas additive and subtractive transfer functions can transmit much or all of it. Modulatory transfer functions transmit all the information in the drive given that it is strong enough, including that which is unique to the drive, whereas divisive and multiplicative transfer functions transmit little or no  information unique to either input, and predominantly transmit synergistic information. Divisive interactions transmit little or no input information under conditions where the amplifying effects of modulation are strongest, i.e. where drive is present but weak and modulatory inputs are strong. Modulatory transfer functions show strong asymmetries between driving and modulatory inputs, as they were designed to do, and this clearly contrasts them with additive and multiplicative functions.

Modulatory, additive, and subtractive transfer functions can have equivalent or similar effects under special input conditions, however, such as when the modulatory input is very weak. Many empirically observed transfer functions that seem linear may therefore seem so only because the modulatory input was weak or absent, as it is in many physiological and cognitive laboratory experiments.

The decompositions obtained when the inputs were continuous were much the same as when they were binary. Thus the present results are likely to be relevant to a wide variety of empirical findings. The main difference between continuous and binary inputs is that in the continuous case more output information was unrelated to the input, but that is to be expected given that only binary outputs were studied here. It will therefore be important to extend these studies so that they include continuous outputs as well as continuous inputs. Even when restricted to binary or discrete outputs the current findings can still be compared with much empirical data, however, because much of neuroscience concerns binary decisions such as spike or no-spike, burst or no burst, excitation or inhibition. Furthermore, as the decompositions presented here have been concerned with a single local processor, they can be compared with the discrete responses to which subjects are restricted in many psychophysical and cognitive experiments. Where this has already been done conditions were observed in which behaviour could be explained by contextual modulation but not by the arithmetic operators~\cite{KIDP}

\subsection{There are various forms of contextual modulation}
All four modulatory interactions met all or most of  the criteria for contextual modulation, and the differences between them were not great. Decompositions of the modulatory interactions observed in the psychophysical experiments performed in the Dering lab~\cite{KIDP} show them to be more similar to our original modulatory interaction $M_1$ than to the others investigated here. It would therefore be worth searching for empirical paradigms that provide evidence for forms of modulation other than $M_1$. If they can be found, which seems likely, then that would further demonstrate the usefulness of such decompositions.

The results reported here can be seen as pure cases of drive and contextual modulation. We have no grounds for supposing that intermediate cases have no functional utility in neural systems, however. On the contrary, it seems probable that intermediate cases combining some properties of drive and modulation in other ways will be found in biological neural systems. 
 
\subsection{Is divisive normalization a form of contextual modulation?}
It has been argued that divisive normalization is a form of contextual modulation~\cite{PCS}, and experiments showing that the output of layer 2/3 pyramidal cells in visual cortex are amplified when the PV interneurons that inhibit them is optogenetically reduced~\cite{ABCS} were cited as providing direct support for that view. The contrast between modulatory and divisive interactions reported above raises a major difficulty for that view, however. Divisive functions transmit synergetic but not unique components. That clearly contrasts with contextual modulation. Divisive operations clearly compute a function that depends on relations between its two inputs, and it has long been known that receptive fields in neocortex are typically defined by relational rather than by absolute quantities. Divisive normalisation predominantly transmits information about relations between its two inputs as shown by the large synergistic components of the output that it produces.  This implies that divisive normalization plays a major role in determining selective sensitivity. This fits well with other evidence that it is basal and perisomatic inputs that specify selective sensitivity because PV inhibitory interneurons synapse on basal and perisomatic post-synaptic sites, not on the apical sites hypothesized to amplify or attenuate transmission of the driving information~\cite{PCS , WAP}. In short, decompositions of contextual modulation and divisive interactions contradict our previous assumption that divisive normalisation is an example of contextual modulation. These possibilities can be explored by using multivariate mutual information decomposition.

\subsection{Can contextual modulation be defined by changes in the rate at which output strength increases with driving input strength?}
One way to think of modulation is as a process that increases or decreases the slope of the function relating output to input. For example, in~\cite{CAR}  `gain modulation' is defined as a change in the slope of the firing-rate curve, corresponding to a multiplicative or divisive scaling, which is distinct from additive or subtractive shifts. This form of gain-modulation has also been referred to as context-sensitive gain control~\cite{ST, SS, Sal2, PCS, PSilv}. The decompositions reported above now suggest that though the various forms of contextual modulation imply changes in slope, changes in slope do not necessarily imply contextual modulation. Both divisive and multiplicative transfer functions change the slope relating output to input, but both transmit components of input that differ greatly from those transmitted by the forms of contextual modulation studied here. This suggests that contextual amplification and attenuation should not be identified with either non-linearity or with changes in input-output slope. Many selective feature sensitivities can be specified by functions that are non-linear. The scenarios used above were chosen to distinguish the driving input that specifies selective sensitivity from the contextual modulation that amplifies or attenuates transmission of information about the driving input. Those criteria are not equivalent to testing for changes in input-output slope.

\subsection{Is coordinate transformation an example of contextual modulation?}
When summarising evidence that gain modulation is a basic principle of brain function, Salinas and Sejnowski~\cite{SS} argue that it is a nonlinear way in which neurons combine information from two (or more) sources, which may be of sensory, motor, or cognitive origin. They argure that gain modulation is revealed when one input, the modulatory one, affects the gain or the sensitivity of the neuron to the other input, without modifying its selectivity or receptive field properties. Though this is similar to contextual modulation Salinas and Sejnowski demonstrate its role in coordinate transformation, and this shows that there are important differences between gain modulation and contextual modulation. Coordinate transformations compute relations between two inputs so they transmit large synergistic components when both inputs are strong. This clearly contrasts with the contextual modulation studied here, which transmits no synergistic information when both inputs are strong.
 
\subsection{Can contextual modulation be adequately defined by contrasting it with drive?}
Many researchers contrast modulation with drive, as do we. Sherman and Guillery~\cite{SG} have long been associated with this terminology, and have provided much of the most convincing physiological evidence for its usefulness~\cite{LS}. From our present perspective some difficult issues arise for this terminology, however. First, there is as yet no consensus on how to define `drive', and different ways of defining it will have different implications for the contrast with modulation. For example, drive could be defined as input that specifies selectivity or receptive field properties~\cite{SS}, as input that is necessary for information to be transmitted~\cite{PCS}, or even as the information whose transmission is modulated. Second, the decompositions reported here show that both additive and subtractive interactions contribute to output semantics. Though `drive' can be defined to mean driving either up or down, it is more likely to be interpreted as excitatory drive. Third, multiplicative and divisive transfer functions predominantly transmit synergistic information that requires knowledge of both inputs, so both inputs are equally essential. Therefore, if drive is defined as either specifying selectivity or as necessary for information to be transmitted, it would need to include multiplicative and divisive interactions in the `drive', even though well-established terminology describes them as being modulatory. Finally, it seems better to define `contextual modulation' by saying what it is than by saying what it is not. This is what the scenarios studied above were designed to do. One way to clarify these issues would be by applying information decomposition to physiological phenomena such as those on which Sherman and Guillery~\cite{SG} base the distinction between drive and modulation.

\subsection{What forms of modulation occur in neocortex?}
Neurophysiological and biophysical evidence for multiplicative and divisive forms of gain modulation have been extensively discussed in a review of `Neuronal Arithmetic'~\cite{silver}. Contextual modulation, as studied here and by others~\cite{Lamme, Sal1}, was not considered in that review, however. Furthermore, multivariate mutual information decomposition has been developed since 2010, and has not yet been extensively applied to physiological or psychophysical phenomena. Where it has been, as in the contextual effects of flankers in psychophysical studies of visual edge detection~\cite{KIDP}, results are encouraging. We know of no studies decomposing information transmitted at the neuronal or microcircuit level under scenarios equivalent to those used here to identify and distinguish various forms of modulation. Such studies would be of great interest, and we note that decomposition could be applied to various output variables including those that signal the strength, confidence, or salience of outputs. Cases of particular interest will be those comparing the contributions from basal and apical inputs to outputs from layer 5 and layer 2/3 pyramidal neurons of mammalian neocortex, because there is evidence that in some modes of function their apical inputs modulate response to their basal and perisomatic inputs~\cite{PCS, WAP}.

Our expectation is that such investigations will reveal various cases that are intermediate between what we have treated as a dichotomy. Application of multivariate decompositions to phenomena such as the various top-down contextual effects reviewed in~\cite{GS} will be of particular interest because they may show that different cases of the top-down contextual modulation reviewed can have different forms not previously distinguished. Another potential domain of application is that of attention because it is a paradigmatic example of a form of modulation that is not required to define RF selectivity. Application of multivariate mutual information decomposition based on classical Shannon entropy do provide evidence that attentional effects conform to the criteria for contextual modulation used here~\cite{PC}, but we know of no studies applying the new measures of information decomposition to this issue. Another set of issues awaiting exploration concerns application of multivariate mutual information decomposition to the classical neuromodulators, such as those of the cholinergic, adrenergic, and seretonergic systems.  Are their modulatory effects multiplicative, divisive, contextual, intermediate, or none of these? 

Another issue to be resolved concerns the transmission of information about signal strength. In all signal-strength scenarios studied above strength was fixed, so there was no information about that to be transmitted. Signal strength is not fixed in neocortex, however. Indeed, the function of contextual modulation is to ensures that it does vary. Thus, the implication of this difference for the decompositions reported here requires clarification.

Information decomposition could be applied to data from computational models as well as to observed neurobiological data. Furthermore, as noted briefly in the final section, it might also be applied to the development of transfer functions and algorithms for machine learning.

\subsection{Can information decomposition aid the design of algorithms for context-sensitive processing and learning?}
The development of algorithms for context-sensitive processing and learning is important to machine learning because the algorithms that are currently the most successful make no use of contextual modulation. This limits their ability to generalize, resolve ambiguities, and solve new problems in new ways. Development of context-sensitive algorithms is also important to neuroscience because applying putative neurocomputational principles to large real-world data sets provides rigorous tests of their information processing capabilities. If no useful capabilities are found for the forms of contextual modulation studied here then that will greatly weaken claims that they have utility in neocortex. If they are found then that will greatly extend our understanding of what they can do and how they do it. 

The findings reported here could aid the development of such context-sensitive algorithms in at least four ways. First, the above decompositions indicate that the transfer function that most generally provides the distinctive information processing properties of contextual modulation is M1, i.e. $T(r, c) = \tfrac{1}{2} r (1 + \exp{ (r c)})$. Second, they show the information processing properties of some other possible transfer functions that could be used to provide such capabilities. Third, multivariate mutual information decomposition could show how relevant and irrelevant nuisance components of information transmission change during the course of learning. This could be done by extending an information theoretic analysis that shows how the hidden layers of hierarchical nets acquire their capabilities during the course of deep learning [33]. This analysis has shown how abstractions of input information producing the required top-level outputs are acquired at each level of the hierarchy during the course of learning. Being based on classical Shannon measures, however, it analyzed the mutual information between just a pair of variables, i.e. either that between the external input to the lowest level of the hierarchy and the output of any level, or that between the output of any level and that required by the top-level supervisory input. An extended form of this analysis could now be applied to context-sensitive nets trained by algorithms such as those in~\cite{KP2}, but with more components of mutual information being monitored. Such an analysis may help us understand what those learning algorithms and architectures can do and how they do it. Fourth, there are a few other algorithms for big data processing that use some form of context-sensitivity, e.g.~\cite{RMCT}, so it would be useful to perform information decompositions of the outputs from the local processing units in such algorithms to see whether they do or do not have properties that are similar to the forms of contextual modulation studied here.

\vspace{0.5cm}

\begin{flushleft}
{\bf Acknowledgements}
\end{flushleft}

\vspace{-0.3cm}
\noindent These results were produced using the excellent package {\it dit} and we thank Ryan James and colleagues for making this package freely available. 

\noindent We are grateful to our many colleagues and collaborators in Glasgow, Stirling, Germany, Norway, Estonia and Singapore for their contributions to research on these and related issues over several years. 

\noindent The work of WAP is in part supported by funding awarded to Lars Muckli 
from the European Union's Horizon 2020 Research and Innovation Programme under Grant Agreement No. 720270 (HBP SGA1).

\newpage


\end{document}